\documentclass[11pt]{article}
\usepackage{times}
\usepackage{sidecap}
\usepackage{geometry}
\usepackage{hyperref}
\usepackage{graphicx}
\usepackage{caption}
\usepackage{xcolor}
\usepackage{tcolorbox}

\usepackage{natbib,bm}

\usepackage{paralist}

\newcounter{aff}

\DeclareGraphicsExtensions{.jpg,.pdf,.png,.eps}

\bibpunct{(}{)}{;}{a}{}{,}

\usepackage{soul}

\definecolor{darkcyan}{rgb}{0.0, 0.0, 0.7}

\definecolor{boxcolor}{RGB}{193, 229, 255}
\definecolor{figboxcolor}{RGB}{193, 229, 255}
\definecolor{boxcolorgray}{RGB}{227, 227, 227}
\definecolor{darkcyan}{rgb}{0.0, 0.0, 0.7}

\sethlcolor{boxcolorgray}

\def\gtrsim{\lower 2pt \hbox{$\, \buildrel {\scriptstyle >}\over
{\scriptstyle \sim}\,$}}
\def\lesssim{\lower 2pt \hbox{$\, \buildrel {\scriptstyle <}\over
{\scriptstyle \sim}\,$}}

\def\approxlt{\lower.2em\hbox{$\buildrel < \over \sim$}}
\def\approxgt{\lower.2em\hbox{$\buildrel > \over \sim$}}

\def\be{\begin{equation}}
\def\ee{\end{equation}}
\def\bea{\begin{eqnarray*}}
\def\eea{\end{eqnarray*}}

\def\jref@jnl#1{{\rm#1}}
\def\aj{\jref@jnl{AJ}}			
\def\araa{\jref@jnl{ARA\&A}}		
\def\apj{\jref@jnl{ApJ}}			
\def\apjl{\jref@jnl{ApJ}}		
\def\apjs{\jref@jnl{ApJS}}		
\def\ao{\jref@jnl{Appl.~Opt.}}		
\def\apss{\jref@jnl{Ap\&SS}}		
\def\aap{\jref@jnl{A\&A}}		
\def\aapr{\jref@jnl{A\&A~Rev.}}		
\def\aaps{\jref@jnl{A\&AS}}		
\def\azh{\jref@jnl{AZh}}			
\def\baas{\jref@jnl{BAAS}}		
\def\jrasc{\jref@jnl{JRASC}}		
\def\memras{\jref@jnl{MmRAS}}		
\def\mnras{\jref@jnl{MNRAS}}		
\def\pra{\jref@jnl{Phys.~Rev.~A}}	
\def\prb{\jref@jnl{Phys.~Rev.~B}}	
\def\prc{\jref@jnl{Phys.~Rev.~C}}	
\def\prd{\jref@jnl{Phys.~Rev.~D}}	
\def\pre{\jref@jnl{Phys.~Rev.~E}}	
\def\prl{\jref@jnl{Phys.~Rev.~Lett.}}	
\def\pasp{\jref@jnl{PASP}}		
\def\pasj{\jref@jnl{PASJ}}		
\def\qjras{\jref@jnl{QJRAS}}		
\def\skytel{\jref@jnl{S\&T}}		
\def\solphys{\jref@jnl{Sol.~Phys.}}	
\def\sovast{\jref@jnl{Soviet~Ast.}}	
\def\ssr{\jref@jnl{Space~Sci.~Rev.}}	
\def\zap{\jref@jnl{ZAp}}			
\def\nat{\jref@jnl{Nature}}		
\def\iaucirc{\jref@jnl{IAU~Circ.}}
\def\aplett{\jref@jnl{Astrophys.~Lett.}}
\def\apspr{\jref@jnl{Astrophys.~Space~Phys.~Res.}}
\def\bain{\jref@jnl{Bull.~Astron.~Inst.~Netherlands}}
\def\fcp{\jref@jnl{Fund.~Cosmic~Phys.}}
\def\gca{\jref@jnl{Geochim.~Cosmochim.~Acta}}
\def\grl{\jref@jnl{Geophys.~Res.~Lett.}}
\def\jcp{\jref@jnl{J.~Chem.~Phys.}}	
\def\jgr{\jref@jnl{J.~Geophys.~Res.}}	
\def\jqsrt{\jref@jnl{J.~Quant.~Spec.~Radiat.~Transf.}}
\def\memsai{\jref@jnl{Mem.~Soc.~Astron.~Italiana}}
\def\nphysa{\jref@jnl{Nucl.~Phys.~A}}
\def\physrep{\jref@jnl{Phys.~Rep.}}
\def\physscr{\jref@jnl{Phys.~Scr}}
\def\planss{\jref@jnl{Planet.~Space~Sci.}}
\def\procspie{\jref@jnl{Proc.~SPIE}}

\def\ion#1#2{#1$\;${\small\rm\@Roman{#2}}\relax}

\newcount\lecurrentfam
\def\LaTeX{\lecurrentfam=\the\fam \leavevmode L\raise.42ex
\hbox{$\fam\lecurrentfam\scriptstyle\kern-.3em A$}\kern-.15em\TeX}
\def\sizrpt{
(\fontname\the\font): em=\the\fontdimen6\font, ex=\the\fontdimen5\font
\typeout{
(\fontname\the\font): em=\the\fontdimen6\font, ex=\the\fontdimen5\font
}}
\definecolor{dark}{rgb}{ 0.34, 0.09, 0.49}

\usepackage{fancyhdr,url}
\usepackage{transparent}
\usepackage{tablefootnote}
\usepackage{mathpazo}

\usepackage{newtxtext}
\usepackage{newtxmath}

\begin{document}

\noindent
\begin{raggedright}
{\Huge\bfseries Probing ordered magnetic fields in galaxies across cosmic time \\[0.5em]
\Large The promise of sensitive full-polarisation observations with ALMA2040
}\end{raggedright}\\

\noindent

Jianhang Chen$^1$, 
Melanie Kaasinen$^2$,
Enrique Lopez-Rodriguez$^{3}$, 
Wouter Vlemmings$^{4}$,
Andrea Lapi$^{5}$,
Rosita Paladino$^{6,7}$,
Hannah Stacey$^{6}$,
Pierre Cox$^{8}$,
Tom J.~L.~C. Bakx$^{9}$,
Andreas Faisst$^{10}$,
Jacqueline Hodge$^{11}$,
Elisabete da Cunha$^{12,13}$,
Leindert A. Boogaard$^{11}$,
Dieu D. Nguyen$^{14}$,
Adarsh Kumar$^{15}$,
Leo Wang$^{16}$,
Gerg\"o Popping$^{6}$,
Willem de Roo$^{17}$,
Simona Vegetti$^{17}$

\vspace{1em}
{
\footnotesize{
  \noindent $^1$ Max-Planck-Institut für extraterrestrische Physik, Gie\ss{}enbachstra\ss{}e, 85748 ,Garching, Germany,\\
$^2$ Research School of Astronomy and Astrophysics, Australian National University, Canberra, ACT 2611, Australia, \\
$^3$ Department of Physics and Astronomy, University of South Carolina, Columbia, South Carolina, \\
$^4$ Department of Physics and Astronomy, Chalmers University of Technology, SE-412 96 Gothenburg, Sweden, \\
$^5$ SISSA, Via Bonomea 265, 34136 Trieste, Italy,\\
$^6$ European Southern Observatory, Karl-Schwarzschild-Straße 2, D-85748 Garching bei München, Germany, \\
$^7$ INAF, Istituto di Radioastronomia, Italy, \\
$^8$ Sorbonne Université, UPMC Université Paris 6 and CNRS, UMR 7095, Institut d’Astrophysique de Paris, 98bis Boulevard Arago, F-75014 Paris, France, \\
$^9$ Department of Physics and Astronomy, Chalmers University of Technology, SE-439 94 Gothenburg, Sweden, \\
$^{10}$ IPAC, California Institute of Technology, 1200 E. California Blvd. Pasadena, CA 91125, USA, \\
$^{11}$ Leiden Observatory, Leiden University, P.O. Box 9513, 2300 RA Leiden, The Netherlands, \\
$^{12}$ International Centre for Radio Astronomy Research, University of Western Australia, \\
$^{13}$ ICRAR M468, 35 Stirling Hwy, Crawley 6009, Western Australia, \\
$^{14}$ Department of Astronomy, University of Michigan, 1085 South University Avenue, Ann Arbor, MI 48109, USA, \\
$^{15}$ Instituto de Astronomía y Ciencias Planetarias (INCT) Universidad de Atacama, Chile, \\
$^{16}$ Department of Astronomy, Cornell University, Ithaca, NY 14853, USA, \\
$^{17}$ Max Planck Institute for Astrophysics, Karl-Schwarzschild Straße 1, D-85748 Garching bei M\"unchen, Germany, 
}
}

\begin{tcolorbox}[boxsep=3pt,left=0pt,right=0pt,top=1pt,bottom=1pt,colframe=boxcolor,colback=boxcolor]
\justify
\textbf{Abstract:} 
Magnetic fields are dynamically important across many astronomical environments, yet their origin, strength, and evolution over cosmic time remain poorly understood.
Ordered magnetic fields are ubiquitous in the nearby Universe, as revealed by a range of observational techniques, but evidence for their existence at cosmological distances is scarce.
Recent ALMA detections of polarised dust emission indicate that galactic-scale ordered magnetic fields, spanning up to kiloparsec scales, were already present in massive star-forming galaxies at redshifts up to $z\sim5.6$.
This white paper argues that the ALMA2040 concept -- with its nearly tenfold improvement in sensitivity and angular resolution, and full-polarisation capability extended to higher-frequency bands -- will open up the study of cosmic magnetism in the early Universe.
We identify four science drivers: constraining the dynamos that generate galactic-scale ordered fields, measuring the strength of early magnetic fields, connecting magnetic fields to galaxy evolution across cosmic time, and constraining dust grain properties through the polarised spectral energy distribution.
We summarise the corresponding requirements in angular resolution, spectral coverage, and sensitivity, and discuss synergies with next-generation telescopes.

\end{tcolorbox}

\newpage{}

\vspace{-0.7cm}
\section{Scientific context and research frontiers}


\subsection{Theoretical frontier}
Magnetic fields are recognised as a fundamental, dynamically important component of the baryonic Universe.
Their importance cascades across scales: they regulate the collapse and fragmentation of molecular clouds and thereby the star formation rate (SFR) and the stellar initial mass function \citep{Federrath2012, Krumholz2019, Stacy2022}; they couple cosmic rays to the interstellar medium (ISM) and drive galactic winds \citep{Steinwandel2022}; they control the accretion and feedback of supermassive black holes \citep{Mckinney2012, Jiang2019, Begelman2023}; and they thread the gas accretion and outflows of the circumgalactic medium (CGM) \citep{Berlok2019,vandeVoort2021}.
Despite their importance, the origin of magnetic fields and the timescale of their effects on galaxy evolution remain unclear.

Two origins of seed magnetic fields have been proposed: primordial and astrophysical. 
Primordial fields could be generated in the early Universe during inflation or at the electro-weak and quark-hadron phase transitions \citep{Widrow2012, Durrer2013, Subramanian2016, Subramanian2019}. Astrophysical fields, instead, are generated from the differential motion of electrons and ions in curved, density-stratified flows \citep[the Biermann battery;][]{Hanayama2005,Xu2008,Seta2020,Attia2021}.
These seed fields are turbulent and generally weak, ranging from $\sim10^{-9}$ to $10^{-17}$~G, but they can be quickly amplified to micro-Gauss levels within millions of years \citep{Federrath2016,Seta2022}.
Later stretching by both cosmic expansion and astrophysical processes, such as differential rotation, gas flows, and shock compression, can create large-scale ordered magnetic fields.\footnotemark[1]
The ordering processes are thought to take much longer \citep{Arshakian2009,Rodrigues2019}; however, the typical length, strength, and lifetime of these ordered fields remain largely unknown, posing fundamental questions about their importance for the earliest star formation and galaxy evolution in the early Universe.

\footnotetext[1]{
In this white paper, we use `coherent' to refer to large-scale magnetic fields that maintain a well-defined direction (i.e., do not reverse sign) along the relevant path or volume, and `ordered' to refer to the plane-of-sky magnetic field morphology inferred from linear polarisation observations, which is subject to an inherent $180^\circ$ directional ambiguity.
}

\subsection{Observational frontier}

Observations of magnetic fields have been fruitful both in Galactic star-forming environments and in nearby galaxies \citep[see reviews][and the white paper on cosmic magnetism by Girart~J.~M. et al.]{Crutcher2012,Beck2013,Lopez-Rodriguez2022,Pattle2023}, but only a few successful attempts have been reported at a cosmological distance.
The first approach is the \emph{Zeeman} effect, which measures the splitting of spectral lines in the presence of ambient magnetic fields.
The main difficulty of observing the Zeeman effect is finding sufficiently strong lines with large splitting, which is largely limited to the H\,{\sc i} 21-cm line and maser lines \citep{Robishaw2026}.
\citet{Wolfe2008} initially reported the highest-redshift Zeeman signature detection in 21-cm absorption system at $z=0.692$ in front of the luminous quasar 3C\,286, but it was later found to be spurious, subsequently providing an upper limit on the field strength of $B_{\rm los}<17\,\mu$G \citep{Wolfe2011}.
At a slightly lower redshift, \citet{Robishaw2008} measured the Zeeman splitting in (Ultra-)luminous Infrared Galaxies ((U)LIRGs) up to $z\sim0.26$, implying that $\mu$G-level magnetic fields likely pervade the interstellar medium of (U)LIRGs \citep[see also][]{McBride2013,Hou2026}.

The current most successful method is {\it Faraday rotation}, which quantifies the change in polarisation angle as light passes through a magnetised medium.
High-redshift polarised quasars are the best targets, but their total rotation measures do not reveal the origin of the rotation \citep{Oren1995, Kronberg2008}.
One way around this difficulty is to measure differential Faraday rotation.
\citet{Bernet2008} showed statistically that quasars with strong Mg\,{\sc ii} absorption lines are associated with large rotation measures.
Assuming that Mg\,{\sc ii} absorption is associated with the halos of normal galaxies along the quasar sightlines, their results favour the presence of coherent magnetic fields of tens of $\mu$G in galaxies up to $z\sim2$ (see also similar works from \citet{Kronberg2008,Bernet2013,Farnes2014,Farnes2017,Malik2026}, but also counter-arguments from \citet{Lan2020}).
\citet{Mao2017} applied this method to a strongly lensed quasar whose two sightlines pass through different regions of a foreground galaxy; the differing Faraday rotation between the two sightlines indicates $\sim10\,\mu$G in the ISM of the foreground galaxy at $z=0.439$.
Fast radio and gamma-ray bursts have become promising new targets \citep{Mannings2023,Brivio2025}, but a systematic search for magnetic fields in the early Universe has to await the next generation of radio telescopes \citep{Beck2019,Mao2026}.
Both Zeeman splitting and Faraday rotation probe only the coherent field along the line of sight (LOS) and offer no information on its geometry, except in the Milky Way and its close neighbours, where multiple LOS measurements are possible.

\subsection{The observation window opened by ALMA}

\begin{figure}[htb!]
    \centering
    \begin{minipage}[c]{0.68\textwidth}
        \includegraphics[width=\textwidth]{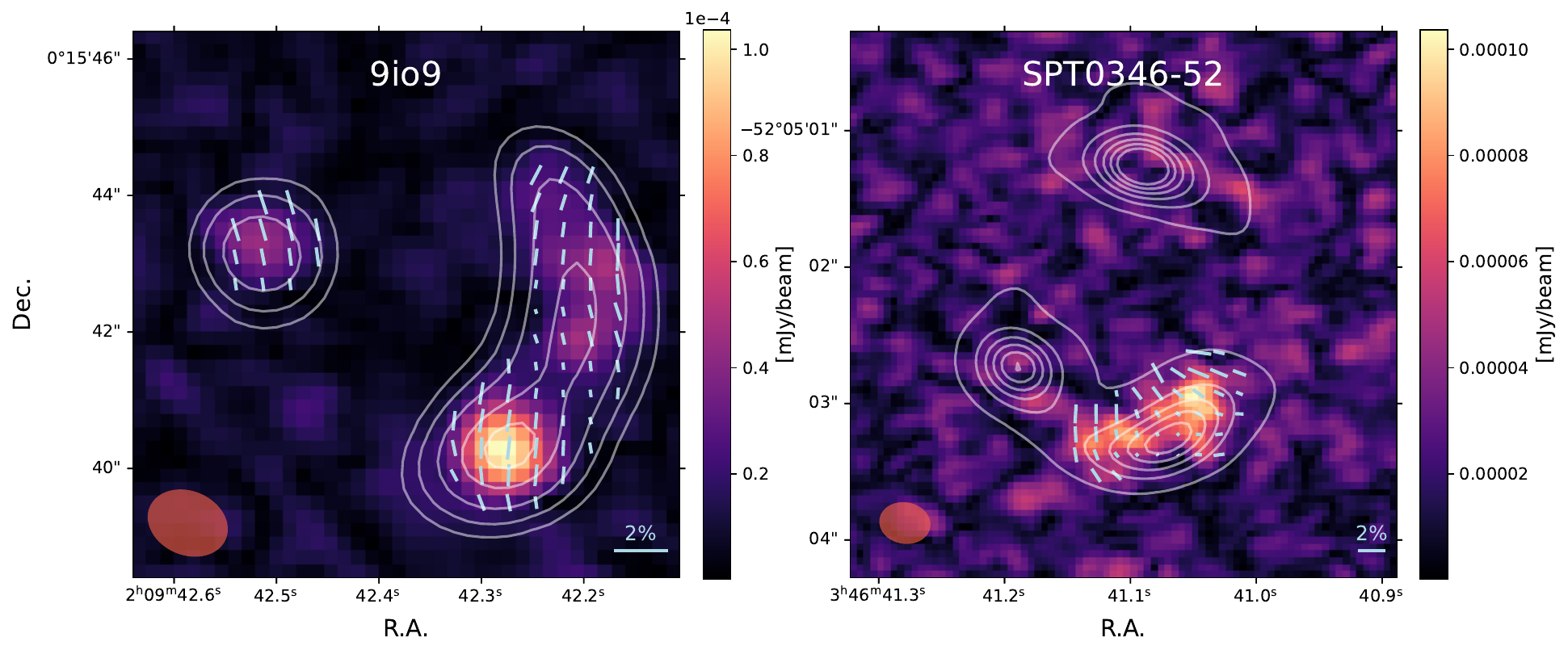}
    \end{minipage}%
    \hfill
    \begin{minipage}[c]{0.30\textwidth}
        \captionof{figure}{The current two successful detections of dust
        polarisation at $z>2$ (\citet{Geach2023, Chen2024}). The images
        show the polarised dust intensity, and the white contour shows
        the total dust intensity. The cyan vectors show the derived
        orientation of the inferred magnetic fields.}
        \label{fig:polarization_detections}
    \end{minipage}
\end{figure}

The full-polarisation capability of ALMA has opened new possibilities for observing the earliest magnetic fields.
For galaxies at $z>1$, ALMA can effectively probe the rest-frame far-infrared (far-IR) emission, where cold dust emission and cold gas spectral lines dominate.
Of particular interest is polarised dust emission, as it has already been widely used in the nearby Universe.
When dust grains are exposed to the interstellar radiation field, the pressure on their irregular surfaces can spin them up via the radiative-torque mechanism \citep{Lazarian2007,Hoang2008}.
The magnetic moments of the rotating paramagnetic grains then quickly align with the external magnetic field via Larmor precession \citep{Andersson2015}.
Once a significant fraction of grains has aligned, they preferentially absorb stellar radiation along the aligned direction, polarising the transmitted starlight, while their own emission also becomes polarised.
This mechanism has become the basis for using polarised dust to probe magnetic fields in star-forming regions and nearby galaxies \citep{Crutcher2012,Lopez-Rodriguez2022}.
The advent of ALMA has made it routine to survey and resolve the dust emission of distant dusty star-forming galaxies \citep[DSFGs;][]{Hodge2020}, and the first searches for polarised dust emission in these galaxies have already yielded promising results \citep[][see also Fig.~\ref{fig:polarization_detections}]{Kade2023, Geach2023, Chen2024}.
Meanwhile, with the help of strong lensing modelling, it is also possible to resolve the field structures in the source plane \citep{Chen2024,deRoo2025}, supporting possible origins with disk formation and galaxy mergers.
Limited by the sensitivity of the current ALMA, these observations primarily focused on the intrinsically hyper-luminous IR galaxies.

Atomic and molecular lines can also become polarised due to the ground-state-alignment (GSA) mechanism \citep{Yan2006,Yan2008} and Goldreich-Kylafis (GK) effect \citep{Goldreich1981}, respectively.
The study and measurement of polarised spectral lines remain less advanced than dust-based observations, primarily due to the unknown polarisation level on galactic scale and the 90$^\circ$--degeneracy uncertainty in the angle between the alignment axis and the magnetic field \citep{Zhang2018,Lankhaar2020}, but these lines offer the potential to reveal the three-dimensional topology of magnetic fields.
The initial attempts to observe molecular lines polarisation have yielded several successes in nearby luminous targets using low-J CO transitions \citep{Li2011,Lopez-Rodriguez2026}, highlighting their potential in the distant Universe.
The first attempt to observe GSA-produced atomic line polarisation with [C\,\textsc{ii}] has yielded only a 3$\sigma$ upper limit on the integrated line at $z>4$ \citep{Kade2023}.
In addition, the spectral lines could originate from regions or gas phases different from the dust continuum, complementing the magnetic fields mapping in the multi-phase, multi-scale ISM.
Future instruments capable of simultaneously resolving both spectral lines and dust continuum in full polarisation could provide more robust evidence of magnetic field structures in early galaxies.

\begin{figure*}[ht]
  \centering
  \includegraphics[width=0.305\textwidth]{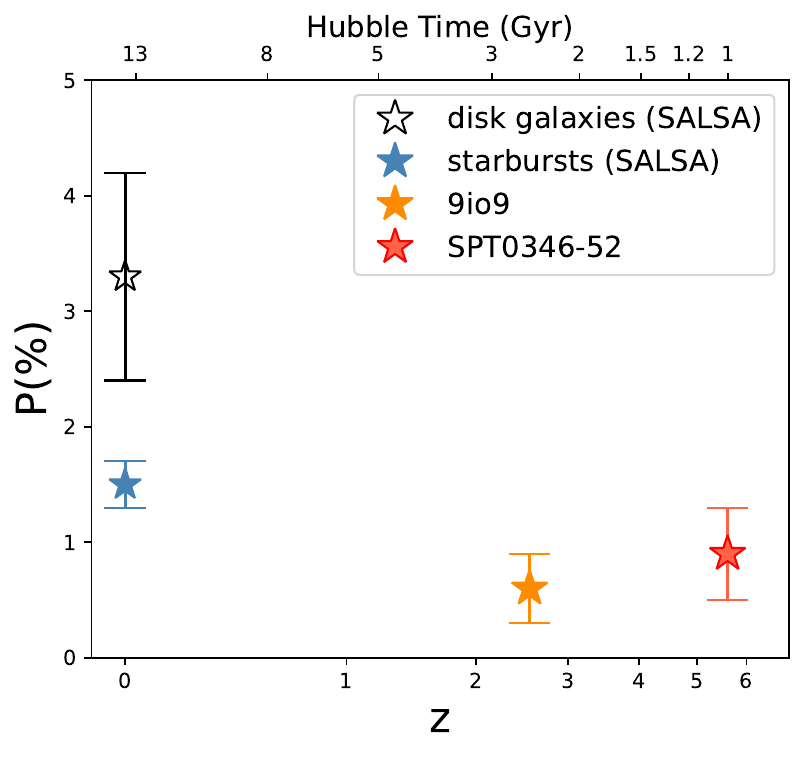}
  \includegraphics[width=0.32\textwidth]{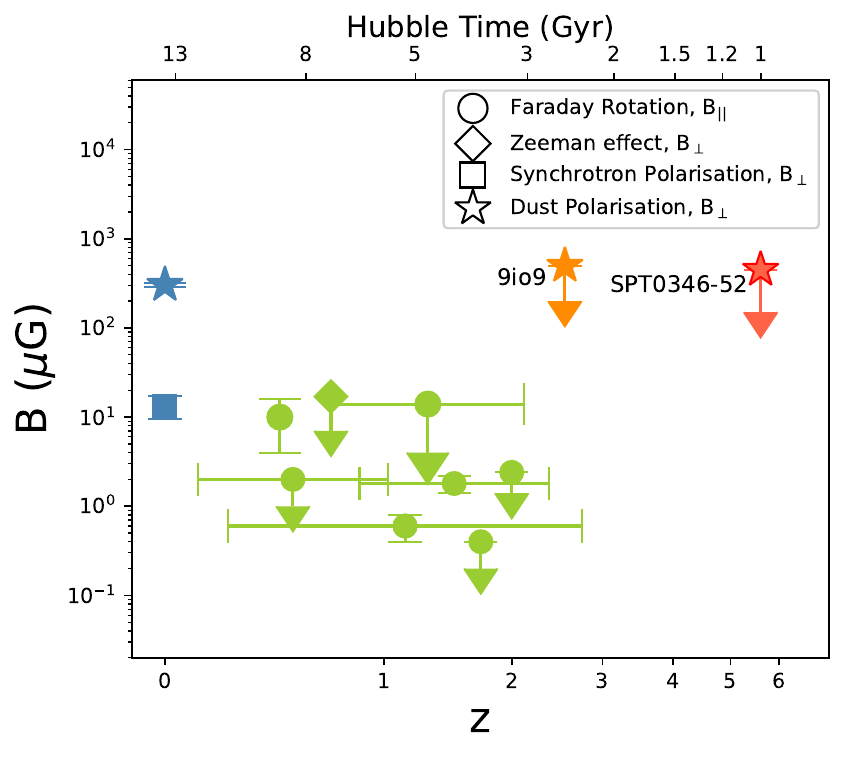}
  \includegraphics[width=0.32\textwidth]{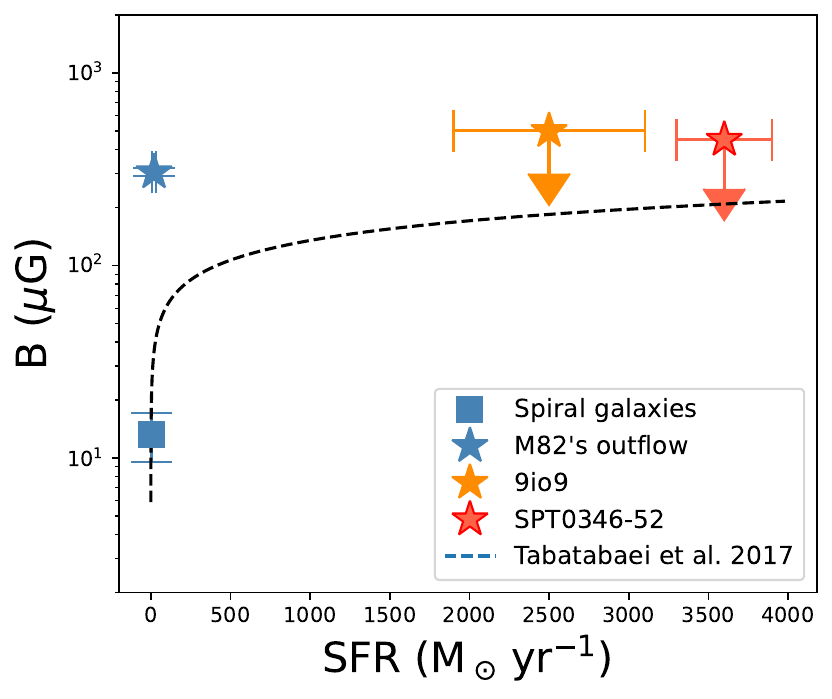}
  \caption{Available measurements of dust polarisation and magnetic fields at different cosmic times and SFR \citep{Chen2024}. \textbf{Left}: Measured polarisation fraction ($P$) as a function of redshift. The measurements of local star-forming and starburst galaxies were adopted from the SALSA Legacy Program \citep{Lopez-Rodriguez2022}. \textbf{Middle}: Summary of the measured magnetic field strength ($B$) in the distant galaxies, collected so far with different methodologies, including the strength of ordered fields measured with polarised dust in the outflow of M82 \citep{Lopez-Rodriguez2021} and the ordered field in local disk galaxies measured with Synchrotron polarisation \citep{Beck2013}, Zeeman splitting \citep{Wolfe2011,Robishaw2008}, and Faraday rotation measurements ($B_\text{los}$) from quasars and AGNs \citep{Mao2017,Bernet2008,Farnes2014,Farnes2017,Lan2020,Pagliotta2026,Malik2026}. \textbf{Right}: Total strength of magnetic fields vs integrated SFR of the host galaxy. For 9io9 and SPT0346-52, their upper limits on the field strength are measured from the equipartition assumption between magnetic fields and gas kinetic energy. The dashed line shows the best-fit model from \citet{Tabatabaei2017}. Current detections of magnetic fields using polarised dust are biased towards extreme starburst galaxies.}
  \label{fig:fpoli_z}
\end{figure*}

\section{ALMA2040}
Current attempts to use ALMA as a cosmic-magnetism probe can target only the brightest sources (see Fig.~\ref{fig:fpoli_z}), which reach the detection limit of ALMA's present capability.
ALMA2040 promises a tenfold increase in both spatial resolution and sensitivity compared with the current ALMA ($\sim8\times$ in spectral line and $\sim5\times$ in continuum relative to the ALMA Wideband Sensitivity Upgrade, WSU), as well as an instantaneous 512~GHz bandwidth coverage across all bands, all of which fully open up the territory of cosmic magnetism in the early Universe.
This section outlines the science drivers that demand the new observing capability and highlights the key requirements for ALMA2040.

\subsection{Science drivers}

\subsubsection{The working dynamos in generating galactic-scale magnetic fields}
\label{subsec:dynamos}

Large-scale dynamos, which stretch small-scale turbulent fields into large-scale coherent fields, are required to generate magnetic fields on kiloparsec scales.
For instance, in rotation-dominated disk galaxies, the differential rotation ($\Omega$-effect) in the disk can stretch and merge twisted fields by the Coriolis force ($\alpha$-effect) to build large-scale toroidal galactic magnetic fields \citep[$\alpha-\Omega$ dynamo,][]{Parker1979,Ruzmaikin1988,Chyzy2008,Mulcahy2017}.
In merging galaxies and in galaxies dominated by outflows, shocks can also drive large-scale stretching and compression, generating ordered fields \citep{Chyzy2004,Lopez-Rodriguez2023b}.
Identifying these dynamos requires resolution-matched observations on the galactic morphology and kinematics.
Meanwhile, the use of polarised dust emission to trace magnetic fields only works if it is aligned by the magnetic radiative torque. 
In real observations, one debate in interpreting polarised dust emission also concerns other non-magnetic mechanisms, such as purely radiative torques \citep{Tazaki2017} and mechanical alignment \citep{Gold1952,Kataoka2019}.
These non-magnetic alignments can align dust grains in dense regions with strong anisotropic radiation or active gas dynamics.
On galactic scales, especially where grain sizes dominate the rest-frame far-IR, magnetic radiative torques are found to be the most effective alignment mechanism \citep{Lopez-Rodriguez2024}, but these tests still need to be performed in distant galaxies, which also requires detailed ISM characterisation.
Compression of an anisotropic turbulent field could also generate polarised emission at scales close to the turbulence \citep{Beck2019}, though this is less likely on kiloparsec scales.  

Current ALMA is already capable of resolving the morphology of gas and dust in distant galaxies and mapping their kinematics using strong emission lines such as CO and [C\,\textsc{ii}].
However, accurate ISM characterisation will significantly benefit from multiple faint FIR spectral lines, such as [N\,\textsc{ii}]122,~205$\mu$m, [N\,\textsc{iii}]57$\mu$m, [O\,\textsc{iii}]52,~88$\mu$m and isotope lines \citep{Doherty2020,Chartab2022}.
The rest frequencies of all these lines usually span multiple ALMA bands, which requires multiple observations and a significant additional time investment.
A future upgrade with multi-band observation can significantly improve the observation efficiency.

\subsubsection{The strength of the early galactic magnetic fields}

With polarisation mapping of both line and continuum, the Davis--Chandrasekhar--Fermi (DCF) method \citep{Davis1951,Chandrasekhar1953} and its derivatives are the most widely used estimators of field strength \citep{Falceta-Goncalves2008,Houde2016}.
DCF relies on the energy equipartition assumption between turbulent kinetic and magnetic energies, which is not universally valid.
Detailed tests against magnetohydrodynamic (MHD) simulations show that DCF is accurate in strong-field regimes but can overestimate the field in weak-field, super-Alfv\'enic situations \citep{Liu2021,Chen2022}.
On the galactic scale, it is also crucial to correct for the contribution of large-scale flows to the velocity dispersion, which would otherwise lead to significantly overestimated field strengths \citep{Lopez-Rodriguez2021a,Guerra2023}.
Accurate ISM conditions, as mentioned in \S\ref{subsec:dynamos}, are required to apply these theoretical corrections.
Meanwhile, the key observational requirement is to resolve the field structures down to the fiducial turbulence scale in the ISM, $\sim$100~parsecs in star-forming galaxies \citep{Elmegreen2004,Haverkorn2008}.

So far, the only way to resolve such a scale and detect a $\sim$1\% polarisation signal is to target strongly lensed sources.
High-frequency observations (above band 8) improve spatial resolution with the current ALMA configuration, but these bands do not offer full polarisation capabilities.
Meanwhile, strong lensing complicates averaging effects because the lensing factor varies across the ISM; ideally, directly resolving the magnetic fields in unlensed sources would provide the most direct measurement, but this would require both better angular resolution and higher sensitivity than ALMA can offer at the moment.


\subsubsection{The connection of galaxy evolution with galactic magnetic fields}

Zoom-in magnetohydrodynamic simulations provide a complementary tool for exploring the impact of early magnetic fields on galaxy evolution.
Simulations with different initial conditions, numerical recipes, and subgrid models can largely reproduce the magnetic fields observed in nearby galaxies \citep{Pakmor2014a,Pakmor2017,Rodrigues2019,Pfrommer2022,Vandenbroucke2021a,Martin-Alvarez2024}, but their predictions for early galaxies differ significantly.
Observations of the early Universe are essential for calibrating these simulations, highlighting the need for unbiased, statistically robust observational constraints.
Two parameters are crucial to this benchmark.
The first is redshift coverage, which sets the range of cosmic time over which galactic magnetic fields can be developed.
In the submm regime, the negative K-correction naturally solves this problem \citep{Hodge2020}, since the submillimetre brightness of DSFGs does not decline significantly with cosmic time; this advantage also ensures that large numbers of galaxies can be observed.
A caveat at high redshift ($z>6$) is that the increasing temperature of the Cosmic Microwave Background (CMB) reduces the contrast of dust emission, leaving only the region with hot dust observable, which also biases the mapping of their magnetic fields.
The second parameter is the FIR luminosity, which serves as a proxy for the SFR of DSFGs.
Along with stellar mass, it allows us to quantitatively characterise the evolutionary stage of DSFGs by their position on the main sequence.

The current sensitivity of ALMA can only reach the brightest lensed galaxies, which represent only a minority of the galaxy population, thereby limiting the parameter space in both cosmic time and galaxy evolutionary stage. 
Expanding the parameter space to include more typical star-forming galaxies, such as main-sequence galaxies, requires a substantial increase in sensitivity.

\subsubsection{Dust polarisation as a probe of grain properties}

Classical two-component dust models assume that the ISM consists of two distinct dust-grain populations, primarily silicate and carbonaceous \citep{Draine2003}.
Such models have been very successful in explaining the extinction curve at optical and near-IR wavelengths.
If these two distinct populations persist in larger grains, the polarisation fraction should increase at longer wavelengths in the far-IR to submillimetre range \citep{Draine2009}.
The two-population model has recently been challenged by far-IR polarimetry, notably the {\it Planck} all-sky polarisation maps, in which the polarisation fraction remains roughly constant across the far-IR to submm window, motivating a single dominant dust population \citep{Hensley2023}.
However, the observed polarisation fraction is not controlled solely by grain composition. 
Temperature mixing, alignment efficiency, magnetic-field geometry along the line of sight and wavelength-dependent depolarisation can all modify the polarised spectral energy distribution (SED). 
Joint forward modelling of multi-band polarimetry, including all these effects, is essential for constraining grain properties and is complementary to general dust models (see the white papers by H. Algera et al. and D. Liu et al.).

The key observational requirement is wide sampling across the frequency range, which is essential for building the SED.
It is the exact limitation of the current ALMA and the WSU upgrade, which requires multiple independent observations to cover each band. 

\subsection{Technical Requirements}

\begin{figure}[ht]
  \centering
  \includegraphics[width=\textwidth]{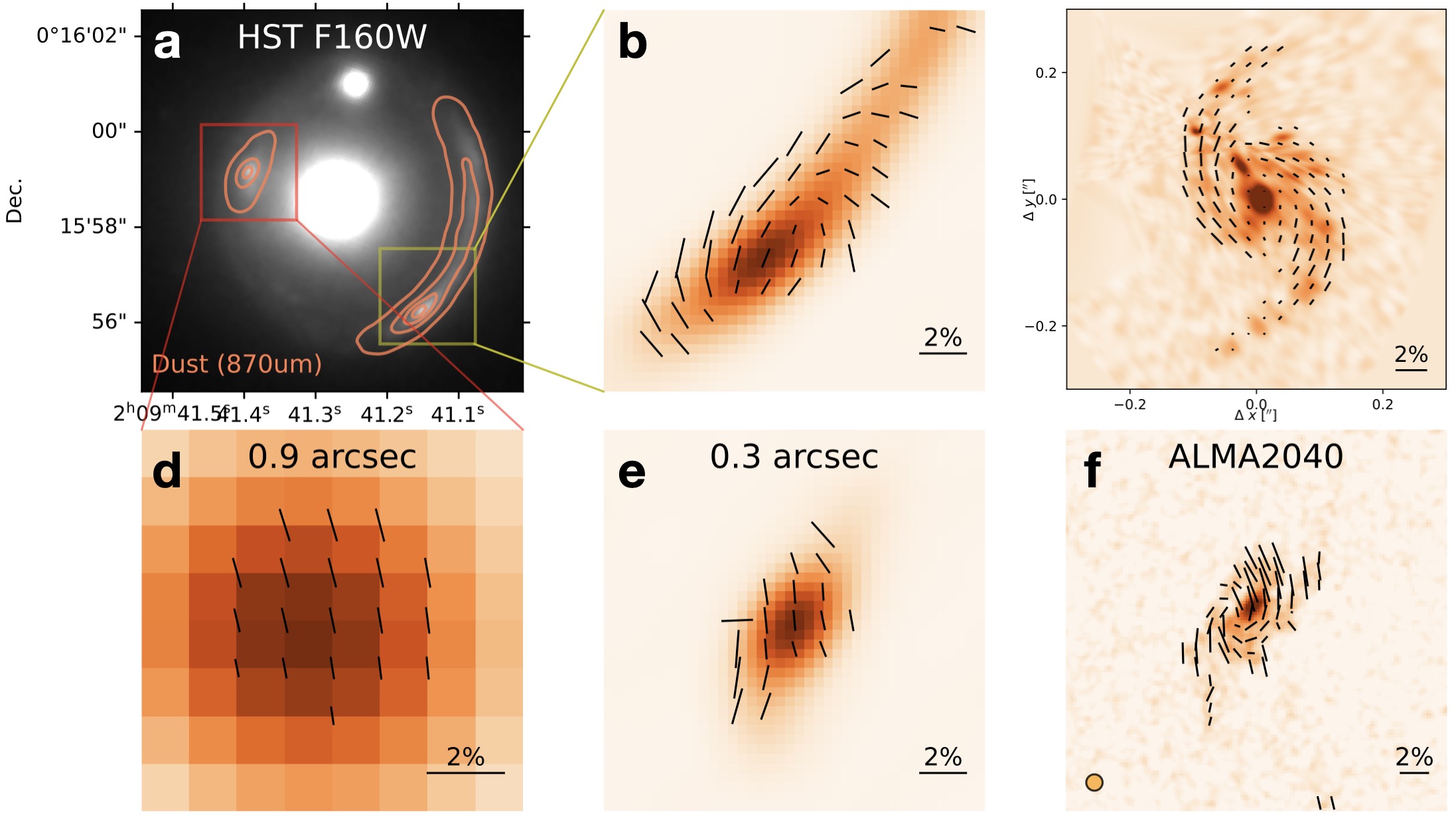}
  \caption{
    Demonstration of mapping magnetic-field structure with future high-resolution observations of unlensed galaxies. The prediction is based on existing observations and modelling of 9io9 at $z\sim2.5$. \textbf{a)}: The HST image of 9io9. The dust continuum observation at 0.3~arcsec resolution is shown as contours. \textbf{b)}: The central part of the right subimage of 9io9 (marked by the yellow square), where the polarisation signal and lens amplification are both strong. \textbf{c)}: The best-fit model of the magnetic fields in the source plane, adopted from \citet{deRoo2025}. \textbf{d \& e)}: The left subimage of 9io9 (marked by the red square), observed with two different spatial resolutions, 0.9 and 0.3~arcsec, respectively. The left subimage shows only mild lensing amplification and distortion, and is a suitable proxy for the unlensed source. \textbf{f)}: The mock observation with intermediate configuration of ALMA2040 (0.06~arcsec). The model image assumes that the ordered magnetic fields align with the two-arm spiral in dust emission. The observations of 9io9 perfectly demonstrate what we can already achieve with current ALMA with 20 hours of telescope time, and what can be done with 1 hour of ALMA2040 observations on a massive unlensed galaxy ($\text{M}_\star\sim10^{11}\text{M}_\odot$) at cosmic noon.
  }
  \label{fig:figs-9io9_simu_image-pdf}
\end{figure}

\subsubsection{Spatial resolution}
The most demanding requirement is to resolve the field structure down to the turbulence scale in a typical main-sequence galaxy at cosmic noon.
Resolving a typical massive star-forming galaxy at cosmic noon, $\text{M}_\star\sim10^{10.5}~\text{M}_\odot$ and SFR~$\sim100~\text{M}_\odot~\text{yr}^{-1}$, down to a turbulence scale of order 50--100~pc with ALMA band 6 or 7, the required resolution is about 5--10~mas, achievable with the extended configuration of the targeted ALMA2040 array.
Extending full-polarisation capability to bands 8--10 could gain another factor of two in spatial resolution.
At lower bands, down to band 4 for instance, where synchrotron and free-free contamination is still negligible, the achievable resolution of about 200~pc at cosmic noon matches the far-IR polarisation mapping of nearby galaxies \citep{Borlaff2023}.
Gravitational lensing remains helpful: it offers the most practical way to observe fainter sources and can directly resolve field structure on cloud scales \citep{Belfiori2025}.
At these highest resolutions, however, the largest recoverable scale will start to miss kpc-scale signals, so a combination with the intermediate and compact configurations will also be needed.

\subsubsection{Spectral coverage}
The current ALMA bands capable of conducting full-polarisation observations are from band 1 to band 7 and can only be observed with one band at a time.
Extending polarisation to higher frequencies and simultaneous sampling of the entire available band are unique advantages of ALMA2040, given the unparalleled site quality.
Higher frequencies also naturally improve spatial resolution: bands 9 and 10 can potentially achieve twice the resolution of canonical band 7 observations, enabling cloud-scale mapping of magnetic fields in some of the viable bright targets. 
A key technical challenge is polarisation calibration at these high frequencies; the band-to-band calibration will be naturally available, but it also requires more systematic studies of polarisation stability and performance.
Lower-frequency bands will also be complementary, both for testing grain-alignment mechanisms and for constraining dust composition.
For instance, self-scattering by millimetre dust grains can also contribute to the continuum polarisation \citep{Kataoka2015}; it mainly dominates the polarisation at the lowest ALMA bands (e.g., bands 1 and 2, $>1$~mm).
At these wavelengths, radio contamination in the ISM becomes significant, requiring full SED modelling to remove it \citep{Belfiori2025}. 
Overall, the targeted 512~GHz instantaneous bandwidth will be the most significant upgrade for improving observation efficiency.

\subsubsection{Sensitivity}
Sensitivity is the key to balancing the desired resolution against the sample size available for statistics.
With ALMA2040, one hour of integration with 128~GHz bandwidth can detect a 1\% polarised signal from an unresolved source of 0.4~mJy, a significant improvement over the current ALMA limit of 24~mJy.
It would also open up the polarised emission from thousands of dusty, star-forming galaxies with $S_{\text{850}\mu\text{m}}>1$~mJy and $P\sim1\%$.
On the survey side, a typical 100-hour large program would resolve sub-kpc field structure in 30 similar galaxies (3 hours each with $>20$ independent beams for star-forming galaxies with $S_{\text{850}\mu\text{m}}>5$~mJy.).
Strong gravitational lensing will continue to be essential for probing even fainter galaxies.
With advanced forward modelling \citep[e.g.,][]{Ndiritu2025}, lensed sources will further complement the diversity of our targets.
In addition to sensitivity, polarisation purity is also critical for detecting faint polarised signals. The current linear polarisation purity of 0.1\% is sufficient for most high-redshift observations, provided that this level of purity can be maintained over the dramatic increase in bandwidth. Circular polarisation purity at a comparable level to that of linear polarisation (e.g. 0.2\%) will be essential for searching for and confirming circular polarisation signals.

%
%
%

\section{Synergies with current and future facilities}

The most complementary synergy comes from next-generation radio telescopes, including both the SKA and the ngVLA.
Radio polarisation observations have long been the workhorse for measuring magnetic field strengths and resolving projected field structures in nearby galaxies \citep{Beck2013}.
Radio polarisation originates from polarised synchrotron radiation emitted by charged particles spiralling along magnetic field lines, and is mainly associated with the diffuse neutral and ionised ISM components \citep{Martin-Alvarez2024}, probing the magnetic fields of the diffuse ISM \citep{Lopez-Rodriguez2023b}.
Dust polarisation, weighted by flux density, traces the projected magnetic fields associated with the molecular clouds.
Both SKA and ngVLA will dramatically improve the sensitivity in the L to C bands (1-10\,GHz), where synchrotron emission can still dominate the radio emission of distant star-forming galaxies.
They will open up full-polarisation synchrotron observations of the same targets accessible to ALMA, providing complementary views of magnetic fields in the multiphase ISM.
Meanwhile, with SKA-mid and ngVLA, the search for Zeeman splitting and Faraday rotation can also be extended to higher redshifts \citep{Robishaw2026,Mao2026}, providing an independent measurement of the coherent field strength along the sightlines of distant galaxies.

Looking ahead to the 2040s, several exciting multi-wavelength facilities will provide the most detailed views of the multiphase ISM.
The Habitable World Observatory \citep[HWO;][]{Dressing2026} will deliver unmatched spatial resolution in the UV, optical, and near-IR, covering both the stellar continuum and spectral lines.
Notably, spectral polarimetry with HWO will enable stellar and spectral-line polarisation observations in the UV and optical \citep[e.g.,] []{Yan2019}, providing independent views of magnetic fields compared with dust emission.
At mid- to far-IR wavelengths, the PRobe far-Infrared Mission for Astrophysics \citep[PRIMA;][]{Moullet2025} will cover the same rest-frame wavelengths as ALMA for distant galaxies, offering a statistical sample of the local Universe to be compared with ALMA.
The detailed mapping of dust polarisation across the local Universe will also provide a valuable reference for interpreting the magnetic fields in the earliest ISM.

\section*{Acknowledgements}
This work was developed as part of the broader ALMA2040 community effort. We acknowledge the Lorentz Center and NOVA for supporting the community workshops and coordination activities that helped initiate and shape the ALMA2040 White Paper process.\\

\bibliographystyle{apj_mod}
{\bibliography{dustpol}}
\end{document}